\documentclass[twocolumn,tighten]{aastex62}

\shorttitle{NuSTAR Observations of 30~Doradus~C}
\shortauthors{Lopez et al.}

\newcommand{\ltsima}{$\; \buildrel < \over \sim \;$}
\newcommand{\simlt}{\lower.5ex\hbox{\ltsima}}

\def\arcdeg{\hbox{$^\circ$}}
\def\arcmin{\hbox{$^\prime$}}
\def\arcsec{\hbox{$^{\prime\prime}$}}

\begin{document}

\title{Evidence of Particle Acceleration in the Superbubble 30~Doradus~C with {\it NuSTAR}}

\correspondingauthor{Laura A. Lopez}
\email{lopez.513@osu.edu}

\author{Laura A. Lopez}
\affil{Department of Astronomy, The Ohio State University, 140 W. 18th Ave., Columbus, Ohio 43210, USA}
\affil{Center for Cosmology and AstroParticle Physics, The Ohio State University, 191 W. Woodruff Ave., Columbus, OH 43210, USA}
\affil{Niels Bohr Institute, University of Copenhagen, Blegdamsvej 17, 2100 Copenhagen, Denmark}

\author{Brian W. Grefenstette}
\affil{Cahill Center for Astrophysics, 1216 E. California Boulevard, California Institute of Technology, Pasadena, CA 91125, USA}

\author{Katie Auchettl}
\affil{Center for Cosmology and AstroParticle Physics, The Ohio State University, 191 W. Woodruff Ave., Columbus, OH 43210, USA}
\affil{Niels Bohr Institute, University of Copenhagen, Blegdamsvej 17, 2100 Copenhagen, Denmark}
\affil{Department of Physics, The Ohio State University, 191 W. Woodruff Ave., Columbus, OH 43210, USA}

\author{Kristin K. Madsen}
\affil{Cahill Center for Astrophysics, 1216 E. California Boulevard, California Institute of Technology, Pasadena, CA 91125, USA}

\author{Daniel Castro}
\affil{Harvard-Smithsonian Center for Astrophysics, 60 Garden Street, Cambridge, MA 02138, USA}

\begin{abstract}
We present evidence of diffuse, non-thermal X-ray emission from the superbubble 30~Doradus~C (30~Dor~C) using hard X-ray images and spectra from {\it NuSTAR} observations. For this analysis, we utilize data from a 200~ks targeted observation of 30~Dor~C as well as 2.8~Ms of serendipitous off-axis observations from the monitoring of nearby SN~1987A. The complete shell of 30~Dor~C is detected up to 20~keV, and the young supernova remnant MCSNR J0536$-$6913 in the southeast of 30~Dor~C is not detected above 8~keV. Additionally, six point sources identified in previous {\it Chandra} and {\it XMM-Newton} investigations have hard X-ray emission coincident with their locations. Joint spectral fits to the {\it NuSTAR} and {\it XMM-Newton} spectra across the 30~Dor~C shell confirm the non-thermal nature of the diffuse emission. Given the best-fit rolloff frequencies of the X-ray spectra, we find maximum electron energies of $\approx70-110$~TeV (assuming a B-field strength of 4$\mu$G), suggesting 30~Dor~C is accelerating particles. Particles are either accelerated via diffusive shock acceleration at locations where the shocks have not stalled behind the H$\alpha$ shell, or cosmic-rays are accelerated through repeated acceleration of low-energy particles via turbulence and magnetohydrodynamic waves in the bubble's interior.


\end{abstract}

\section{Introduction}

OB associations typically have tens of massive stars, and the collective effect of their fast stellar winds and supernovae (SNe) create superbubbles (SBs; e.g., \citealt{maclow88,oey96,yadav17}). SBs are large ($\sim$100~pc) shell-like structures that sweep up material from the surrounding interstellar medium (ISM), producing tenuous cavities filled with hot ($\sim$10$^{6}$~K), shock-heated gas (e.g., \citealt{castor75,weaver77,chu90,rogers14}). Due to the low densities within these bubbles ($n_{\rm ISM} \sim 0.01$~cm$^{-3}$), shock waves travel large distances before substantial deceleration, and thus the timescale of efficient particle acceleration is longer than in the case of individual/isolated supernova remnants (SNRs). Since SBs also have large energy reservoirs, SBs are plausible candidates for sites of cosmic-ray acceleration (e.g., \citealt{bykov92,par04,butt08,ferrand10,bykov14}).  

Observational evidence of particle acceleration in SBs is growing. GeV gamma-rays have been detected by the {\it Fermi} Gamma-ray Space Telescope toward some SBs (e.g., \citealt{abdo10}), and possible detection of non-thermal X-rays from SBs have been reported from multiple sources (e.g., 30 Doradus~C: \citealt{bamba04}; N51D: \citealt{cooper04}; N11: \citealt{maddox09}; IC~131: \citealt{tullman09}; see also the recent review by \citealt{kavanagh20}). However, in some cases, follow-up work failed to find diffuse non-thermal X-rays in these sources, suggesting that the previous findings may be due to inadequate background subtraction or unresolved point sources (e.g., \citealt{yamaguchi10}). Consequently, the detection of diffuse, non-thermal X-rays in SBs remains controversial, and additional observational constraints are necessary to elucidate the role of SBs in the particle acceleration process. 

In this paper, we present hard ($>$3~keV) X-ray images and spectra from {\it NuSTAR} observations of the superbubble 30~Doradus~C (hereafter, 30~Dor~C) in the Large Magellanic Cloud (LMC). 30 Dor~C is a $\approx$95~pc across SB \citep{dunne01} powered by the OB star association LH~90 \citep{lucke70}, with 26 O stars and 7 Wolf-Rayet (WR) stars with ages of 3--7~Myr \citep{testor93}. 

30~Dor~C was first detected in X-rays by the {\it Einstein} Observatory \citep{long81}, and {\it ROSAT} observations resolved the shell-like structure \citep{dunne01}. {\it XMM-Newton} observed 30~Dor~C as its first light image in January 2000 \citep{dennerl01}, revealing hard X-ray emission up to 5~keV. Subsequently, \cite{bamba04} searched for synchrotron emission in archival {\it Chandra} and {\it XMM-Newton} data and found that the southeast of 30 Dor~C has enhanced thermal and line emission, while the emission from the rest of the shell likely arises from non-thermal processes. The spectra in the latter locations were adequately fit by a synchrotron model of an exponentially cut off power-law distribution of electrons \citep{reynolds98}. \cite{yamaguchi09} confirmed the \cite{bamba04} results using {\it Suzaku} data, and \cite{kavanagh15} revisited the {\it XMM-Newton} data and presented evidence for a young SNR, MCSNR J0536$-$6913, in the southeast of 30~Dor~C based on enhanced abundances of intermediate-mass elements there. \cite{babazaki18} considered the {\it XMM-Newton} data and found that all locations in 30~Dor~C have non-thermal X-ray emission, consistent with the previous results of \cite{bamba04} and \cite{kavanagh15}. \cite{sano17} demonstrated that non-thermal X-rays were particularly enhanced in the western shell of 30~Dor~C where they detected several molecular clouds; these authors interpreted the result as evidence that magnetic-field amplification resulted from shock-cloud interaction there.

The High Energy Stereoscopic System (HESS) detection of 30~Dor~C in TeV gamma-rays is additional evidence of particle acceleration \citep{HESS15}. Recently, \cite{kavanagh18} exploited new {\it Chandra} observations of 30~Dor~C to estimate the B-field in the post-shock region using radial profiles around the synchrotron-dominated shell. They found that the filament widths indicated a $B \lesssim 40$~$\mu$G, which is consistent with a leptonic origin of the TeV emission. \cite{kavanagh18} also showed an anti-correlation between the H$\alpha$ and X-ray synchrotron emission in 30~Dor~C, as has been observed in several SNRs (e.g., RCW~86: \citealt{yamaguchi16}). They measured an expansion velocity of the H$\alpha$ shell as $\lesssim$100~km~s$^{-1}$, yet shock velocities of $\gtrsim$1000~km~s$^{-1}$ are necessary to produce the observed X-ray synchrotron emission \citep{bell04}. They interpreted this result as evidence that the non-thermal X-rays originate from locations where the shock continues to expand rapidly in the gaps of the H$\alpha$ shell, while the shock has slowed elsewhere when it encountered that shell.

{\it NuSTAR}, the first satellite to focus hard X-rays at energies 3--79~keV \citep{harrison13}, has observed 30~Dor~C serendipitously fifteen times during its extensive monitoring of SN~1987A $\sim$5\arcmin\ away \citep{boggs15}. Additionally, we obtained two targeted observations of 30~Dor~C totaling 203~ks in late 2015 as part of the guest observer program. The primary scientific objective of our work was to detect and localize hard X-rays from 30~Dor~C and to characterize particle acceleration properties using spectroscopic analysis. 

The paper is organized as follows. In Section~\ref{sec:data}, we outline the {\it NuSTAR} observations and data of 30~Dor~C as well as complementary X-ray data from {\it XMM-Newton} and {\it Chandra}. In Section~\ref{sec:results}, we present the results, including the hard X-ray images of 30~Dor~C (in Section~\ref{sec:images}) as well as spatially-resolved spectral modeling of the shell (in Section~\ref{sec:spectra}). In Section~\ref{sec:discuss}, we discuss the implications regarding SBs and their particle acceleration processes, and in Section~\ref{sec:conclusions}, we summarize our conclusions.

\begin{deluxetable}{lcrcc}
\tablecolumns{5} \renewcommand{\arraystretch}{0.8}
\tablewidth{0pt} \tablecaption{{\it NuSTAR} Observation Log \label{table:obslog}} 
\tablehead{\colhead{\#} & \colhead{ObsID} & \colhead{Exposure} & \colhead{UT Start Date} & \colhead{Off-Axis Angle\tablenotemark{b}}}
\startdata
1 & 40001014002 & 68~ks & 2012-09-07 & 3.4\arcmin \\
2 & 40001014003 & 136~ks & 2012-09-08 & 1.9\arcmin \\
3 & 40001014004 & 199~ks & 2012-09-11 & 2.4\arcmin \\
4 & 40001014006 & 54~ks & 2012-10-20 & 3.5\arcmin \\
5 & 40001014007 & 200~ks & 2012-10-21 & 4.0\arcmin \\
6 & 40001014009 & 28~ks & 2012-12-12 & 5.3\arcmin \\
7 & 40001014010 & 186~ks & 2012-12-12 & 3.1\arcmin \\
8 & 40001014012 & 19~ks & 2013-06-28 & 3.1\arcmin \\
9 & 40001014013 & 473~ks & 2013-06-29 & 3.7\arcmin \\
10 & 40001014015 & 97~ks & 2014-04-21 & 4.2\arcmin \\
11 & 40001014016 & 432~ks & 2014-04-22 & 5.4\arcmin \\
12 & 40001014018 & 200~ks & 2014-06-15 & 4.4\arcmin \\
13 & 40001014020 & 275~ks & 2014-06-19 & 4.0\arcmin \\
14 & 40001014022 & 48~ks & 2014-08-01 & 2.4\arcmin \\
15 & 40001014023 & 427~ks & 2014-08-01 & 2.9\arcmin \\
16\tablenotemark{a} & 40101015002 & 167~ks & 2015-09-03 & 1.3\arcmin \\
17\tablenotemark{a} & 40101015004 & 26~ks & 2015-10-18 & 0.4\arcmin \\
\enddata
\tablenotetext{a}{Observations \#16 and \#17 are the targeted observations of 30~Dor~C.}
\tablenotetext{b}{Off-axis angle from the aimpoint to position of 30~Dor~C in the Simbad database \citep{wenger00}, taken from \cite{filipovic95}.}
\end{deluxetable}

\begin{figure*}
\begin{center}
\includegraphics[width=\textwidth]{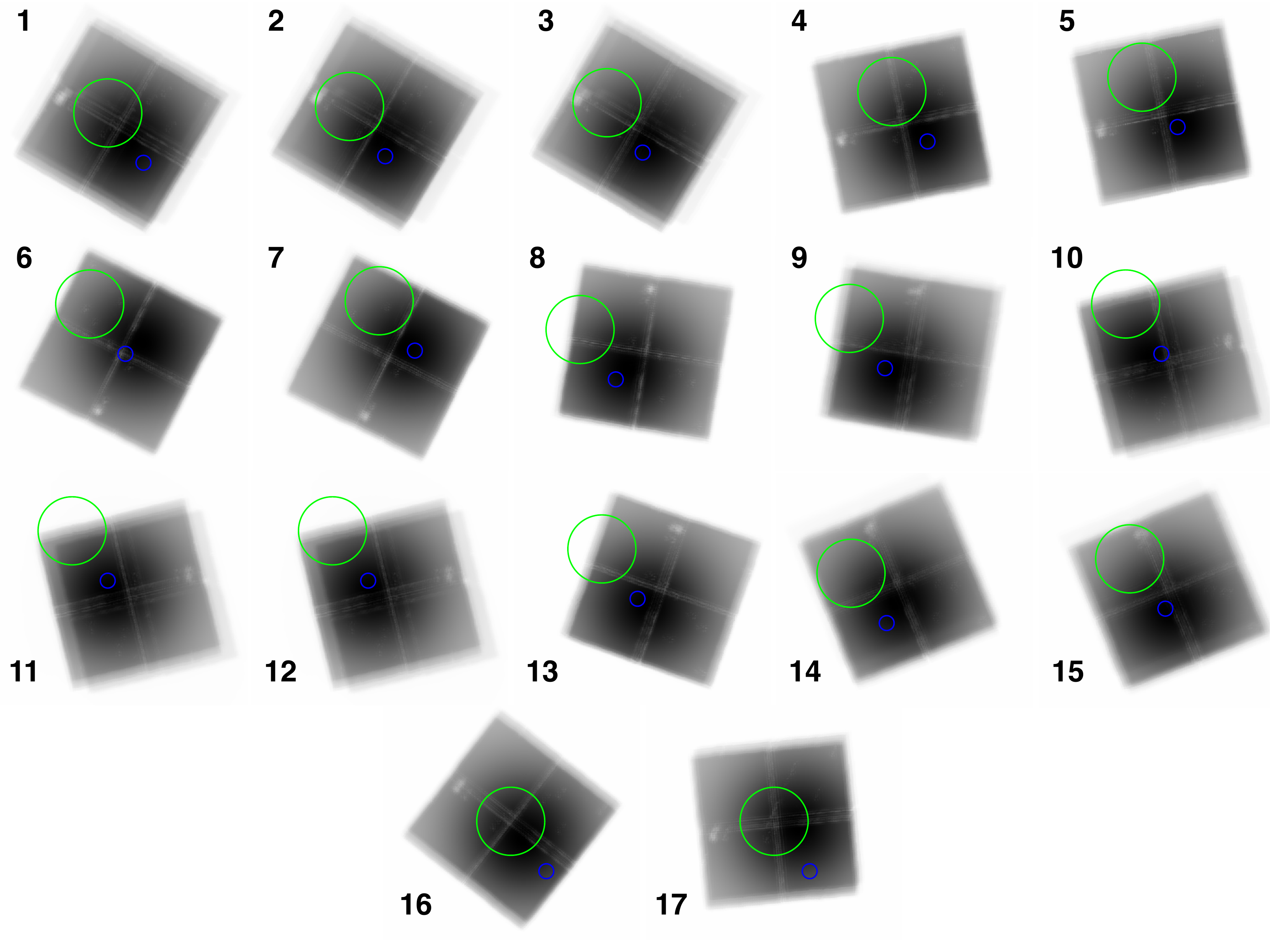}
\end{center}
\caption{Exposure maps with vignetting correction at 10~keV from all seventeen FPMA {\it NuSTAR} observations (with a field of view of 12\arcmin) of the SN~1987A/30~Doradus C region, with circles denoting the location of SN~1987A (blue circles) and of the 30 Dor C shell (green circles; 3\arcmin\ in radius) from {\it Chandra} data. Observations are numbered as in Table~\ref{table:obslog} which gives the relevant information about each field; observations \#16 and \#17 were our targeted observations of 30~Dor~C. North is up, and East is left.}
 \label{fig:expmaps}
\end{figure*}

\section{Observations and Data Analysis} \label{sec:data}

\subsection{{\it NuSTAR} Data} \label{sec:nustardata}

As mentioned above, 30~Dor~C has been observed by {\it NuSTAR} seventeen times: two targeted observations in September and October 2015 as well as fifteen observations during monitoring of SN~1987A. Details of the seventeen observations are outlined in Table~\ref{table:obslog}, and Figure~\ref{fig:expmaps} shows the positions on the detectors of the 30~Dor~C shell and SN~1987A in each observation.

We reduced the data using the {\it NuSTAR} Data Analysis Software (NuSTARDAS) Version 1.8.0 and {\it NuSTAR} CALDB Version 20170817. We performed the standard pipeline data processing with {\it nupipeline}, with the \texttt{saamode=STRICT} to identify the South Atlantic Anomaly (SAA) passages. Using the resulting cleaned event files, we produced images of different energy bands using the FTOOL {\it xselect} and generated associated exposure maps using {\it nuexpomap} with vignetting correction at 10~keV. As 30~Dor~C is an extended source, we opted to model the background and produce synthetic, energy-dependent background images for background subtraction. We followed the procedure outlined by \cite{wik14} to estimate background components and their spatial distribution. Subsequently, we combined the vignetting- and exposure-corrected FPMA and FPMB images from all epochs using {\it ximage}. 

\begin{figure*}
\begin{center}
\includegraphics[width=\textwidth]{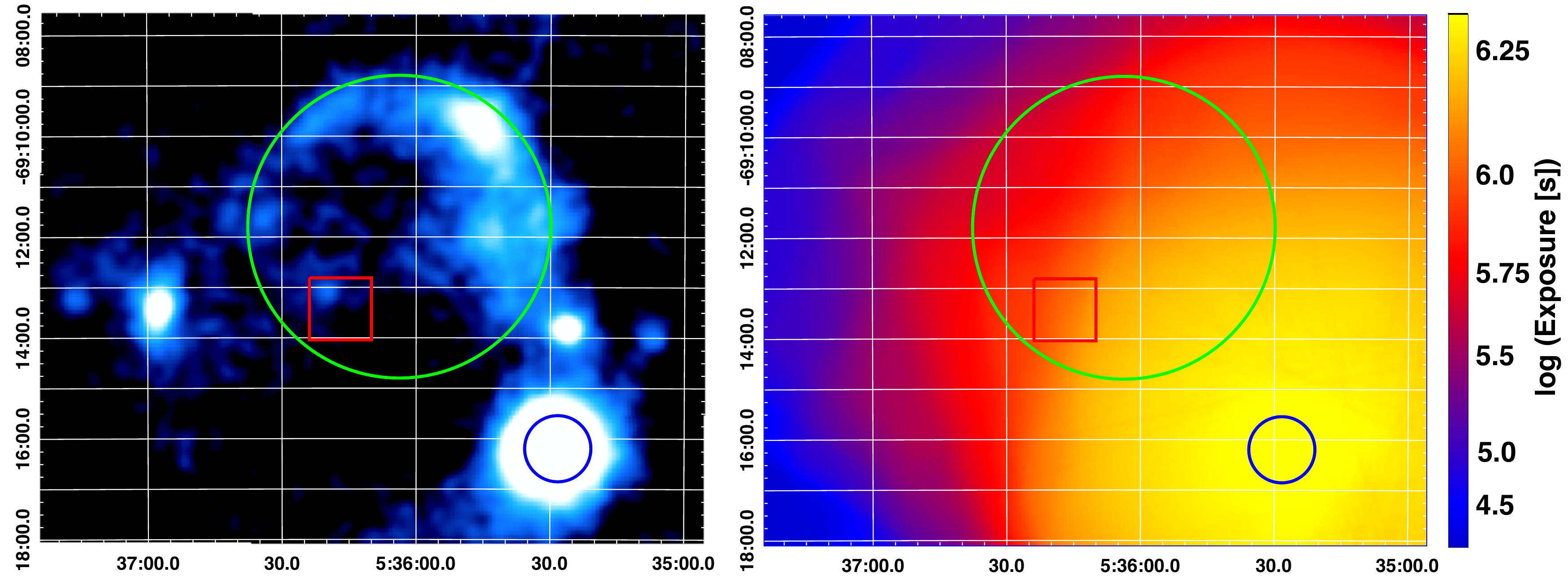}
\end{center}
\vspace{-5mm}
\caption{{\it Left}: Background-subtracted {\it NuSTAR} image of the $3-20$~keV band. To produce this image, we have merged observations that fully image the shell of 30~Dor~C (observations \#1--5, 14, 16--17 in Table~\ref{table:obslog}. The green circle (3\arcmin\ in radius; the same as in Figure~\ref{fig:expmaps}) denotes the location of the 30~Dor~C shell from {\it Chandra} images. The red box marks the position of the MCSNR~J0536$-$6913, and SN~1987A is marked with a blue circle. {\it Right}: The combined exposure map of the same area shown in the left panel. The southwestern part of the shell has the longest effective exposure (of $\sim$1.8~Ms, and the northeastern section has the shortest effective exposure (of $\sim$0.4~Ms).}
 \label{fig:fullfield}
\end{figure*}

The combined images were deconvolved by the on-axis {\it NuSTAR} point-spread function (PSF) using the {\it max\_likelihood} AstroLib IDL routine\footnote{See https://github.com/wlandsman/IDLAstro}. The script employs Lucy-Richardson deconvolution, an iterative procedure to derive the maximum likelihood solution. We set the maximum number of iterations to 20, as more iterations did not lead to any significant changes in the resulting images. We note that this routine assumes that the data can be characterized by a Poisson distribution, but background subtraction causes the images not to follow strictly a Poisson distribution. Thus, the deconvolved images are presented for qualitative purposes only, and we do not use them for any quantitative results. 

\begin{figure*}
\begin{center}
\includegraphics[width=\textwidth]{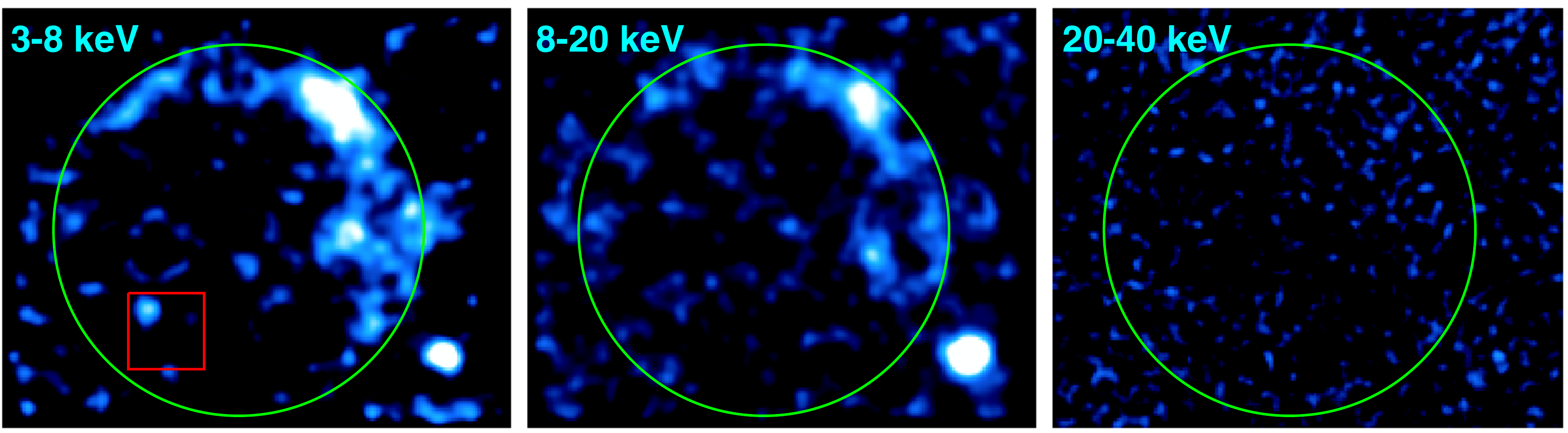}
\end{center}
\vspace{-5mm}
\caption{Deconvolved, background-subtracted {\it NuSTAR} images of 30~Dor~C in three energy bands: $3-8$ keV (left), $8-20$ keV (middle), and $20-40$ keV (right). To produce these images, we have merged observations that fully imaged the shell of 30~Dor~C (observations \#1--5, 14, 16--17 in Table~\ref{table:obslog}). The red box denotes the location of MCSNR J0536$-$6913, and the green circle (same as in Figure~\ref{fig:expmaps}; 3\arcmin\ in radius) marks the position of the 30 Dor C shell to guide the eye. North is up, and East is left.}
 \label{fig:images}
\end{figure*}

We performed a spatially-resolved spectroscopic analysis by extracting and modeling spectra from several locations in 30~Dor~C. Using the {\it nuproducts} FTOOL, we extracted source spectra and produced ancillary response files (ARFs) and redistribution matrix files (RMFs) from each observation and both the A and B modules (17 ObsIDs $\times$ 2 modules $=$ 34 spectra per region). We employed the {\it nuskybgd} routines\footnote{https://github.com/NuSTAR/nuskybgd} (presented in detail in \citealt{wik14}) to simulate associated background spectra. 

The FPMB data had substantial stray-light contamination to the north and southwest of 30~Dor~C from three sources within 5$^{\circ}$ of the object: LMC X--1 (0.629$^{\circ}$ away), 2MASX~J05052442$-$6734358 (3.252$^{\circ}$ away), and IGR~J05007$-$7047 (3.39$^{\circ}$ away). Thus, the area available for background regions was limited to the east and northwest in the FPMB data, while annuli around 30~Dor~C could be employed as background regions for the FPMA data. Consequently, the background subtraction in the FPMA data is more reliable as it samples and adequately accounts for the spatial variation of the background across the source. 

\subsection{{\it XMM-Newton} Data} \label{sec:xmmdata}

To supplement the {\it NuSTAR} data, we also downloaded eleven 30~Dor~C observations from the {\it XMM-Newton} Science Archive. Ten of these observations from 2000--2012 were presented in \cite{kavanagh15}, and the eleventh observation (ObsID 0743790101) was obtained after the submission of that work. We employed the {\it XMM-Newton} Science Analysis System (SAS) Version 15.0.0 and up-to-date calibration files to reduce and analyze this data. All event files were filtered to remove flagged events and periods of high background or photon flare contamination, as identified based on count-rate histograms of the $>$10~keV band. The effective exposure times of the MOS1, MOS2, and pn detectors in the ObsID 0743790101 observation was 77.2~ks, 77.3~ks, and 66.6~ks, respectively. Thus, the net exposures, when combined with the observations analyzed in \cite{kavanagh15}, were 633~ks for MOS1, 692~ks for MOS2, and 487~ks for the pn detector.

{\it XMM-Newton} spectra were extracted using the SAS command \texttt{evselect} on the cleaned, vignetting-corrected event files from the three EPIC cameras. For each region, we produced ARFs and RMFs using the tasks \texttt{arfgen} and \texttt{rmfgen}, and each spectrum was grouped to a minimum of 25 counts per bin. Given that the {\it XMM-Newton} observations spanned 12~years, we opted not to combine the spectra and instead fit the spectra simultaneously.

Given that 30~Dor~C is an extended object and that the background of {\it XMM-Newton} varies across the detectors, we opted to model the background rather than subtract it in our spatially-resolved spectroscopic analysis. We extracted background spectra from a circular, 0.85\arcmin\ radius region northwest of 30~Dor~C, and we modeled the instrumental and astrophysical X-ray background (AXB) contributions similarly to \cite{maggi16} who analyzed {\it XMM-Newton} observations of 51 LMC SNRs. The former components account for the particle-induced background and the soft-proton contamination (see Appendix~A of \citealt{maggi16}), and the latter reflect the emission from the Local Hot Bubble, the Galactic halo, and unresolved background active galactic nuclei (AGN; \citealt{snowden08}). We included these best-fit background components in our models of each {\it XMM-Newton} spectrum from 30~Dor~C to assess accurately the emission from the source regions.

\subsection{{\it Chandra} Data}

To aid in the identification of point sources, we analyzed the available {\it Chandra} data on 30~Dor~C to localize the regions of bright {\it NuSTAR} emission. {\it Chandra}'s Advanced CCD Imaging Spectrometer (ACIS) has imaged SN~1987A repeatedly, and several of those programs included serendipitous coverage of 30~Dor~C. Using the {\it Chandra} archive, we identified nine archival ACIS observations with partial coverage of 30~Dor~C (ObsIDs 1044, 1967, 2831, 2832, 3829, 3830, 4614, 17904, 19925). In addition, 30~Dor~C was observed in December 2018 (ObsIDs 20339, 21949, 22006; PI: Lopez) for $\approx$91~ks and will be presented in detail in future work. We have reprocessed all of the archival and new observations and produced composite, exposure-corrected image of the broad-band (0.5--7.0~keV) using the {\it flux\_image} command in the {\it Chandra} Interactive Analysis of Obserations ({\sc ciao}) software Version 4.7. 

\begin{figure}
\begin{center}
\includegraphics[width=\columnwidth]{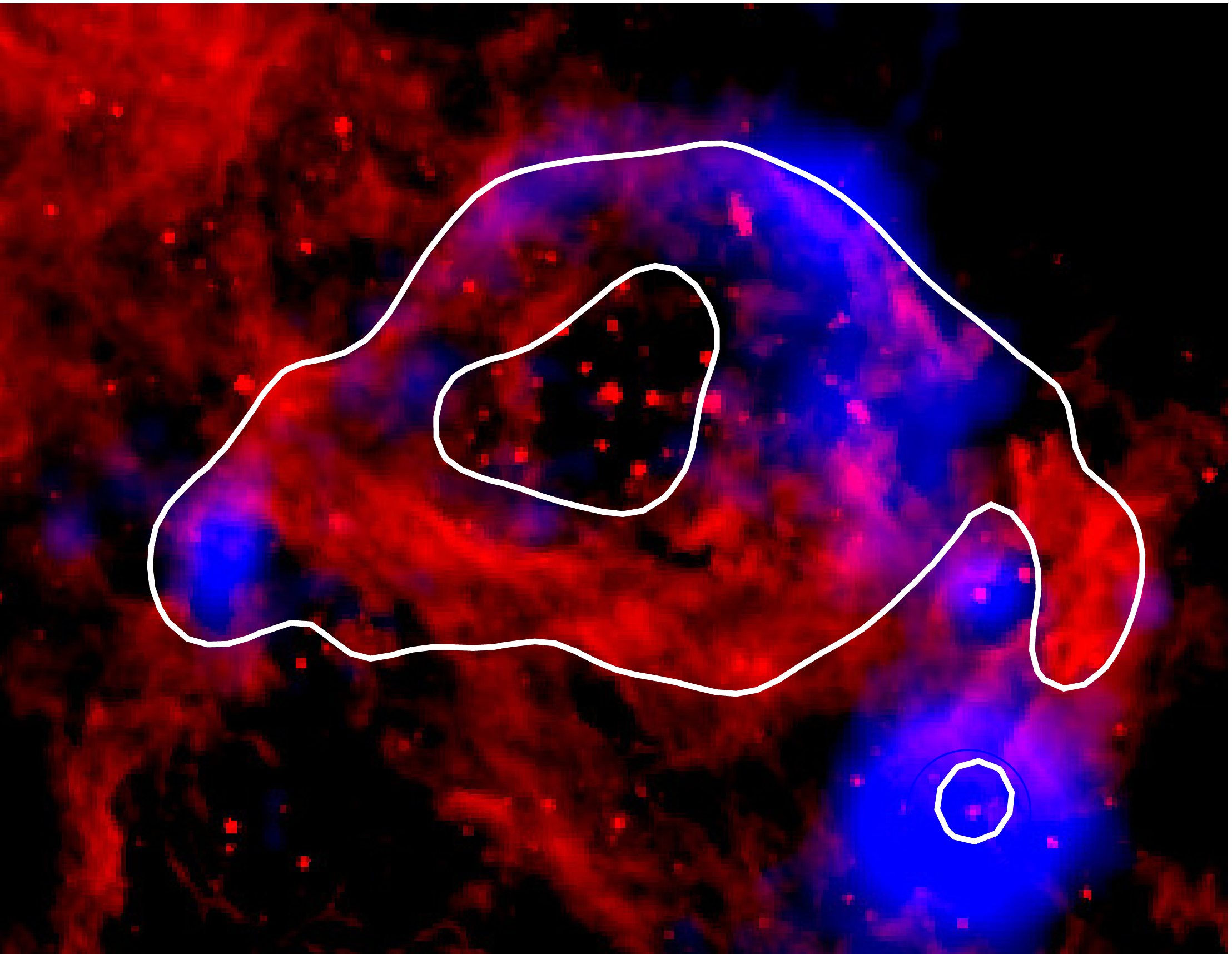}
\end{center}
\vspace{-5mm}
\caption{Two-color image of 30~Dor~C, with H$\alpha$ in red (from \citealt{smith98}) and $3-20$~keV deconvolved, background-subtracted {\it NuSTAR} data in blue (the same X-ray data shown in Figure~\ref{fig:fullfield}). The white contours represent the 1.4~GHz morphology (from \citealt{hughes07}). The complete shell of 30~Dor~C is evident in the radio band. The H$\alpha$ emission particularly correlates with the hard X-rays in the northwest part of the shell where the former emission is particularly narrow. North is up, and East is left.}
\label{fig:threecolor}
\end{figure}

\section{Results} \label{sec:results}

\subsection{Images} \label{sec:images}

\subsubsection{Diffuse Emission} 

Figure~\ref{fig:fullfield} shows the background-subtracted $3-20$ keV image of 30~Dor~C and the nearby sources, including SN~1987A, and Figure~\ref{fig:images} presents the deconvolved, background-subtracted {\it NuSTAR} images of the 30~Dor~C shell in several energy bands. To produce these images, we have merged the eight {\it NuSTAR} observations (observations labeled \#1--5, 14, 16--17 in Figure~\ref{fig:expmaps}) that cover the full extent of the 30~Dor~C shell. {\it NuSTAR} detects the entire rim of 30~Dor~C studied previously in X-rays with {\it Chandra}, {\it XMM-Newton}, and {\it Suzaku}. The brightest emission is in the northwest, where the non-thermal X-rays are detected up to $\sim$20~keV. No emission from any region of 30~Dor~C is detected with significance above 20~keV. In the southeast, we find hard X-rays up to $\sim$8~keV where \cite{kavanagh15} identified a young SNR dubbed MCSNR~J0536$-$6913, based on enhanced abundances of intermediate-mass elements there.

\begin{figure}
\begin{center}
\includegraphics[width=\columnwidth]{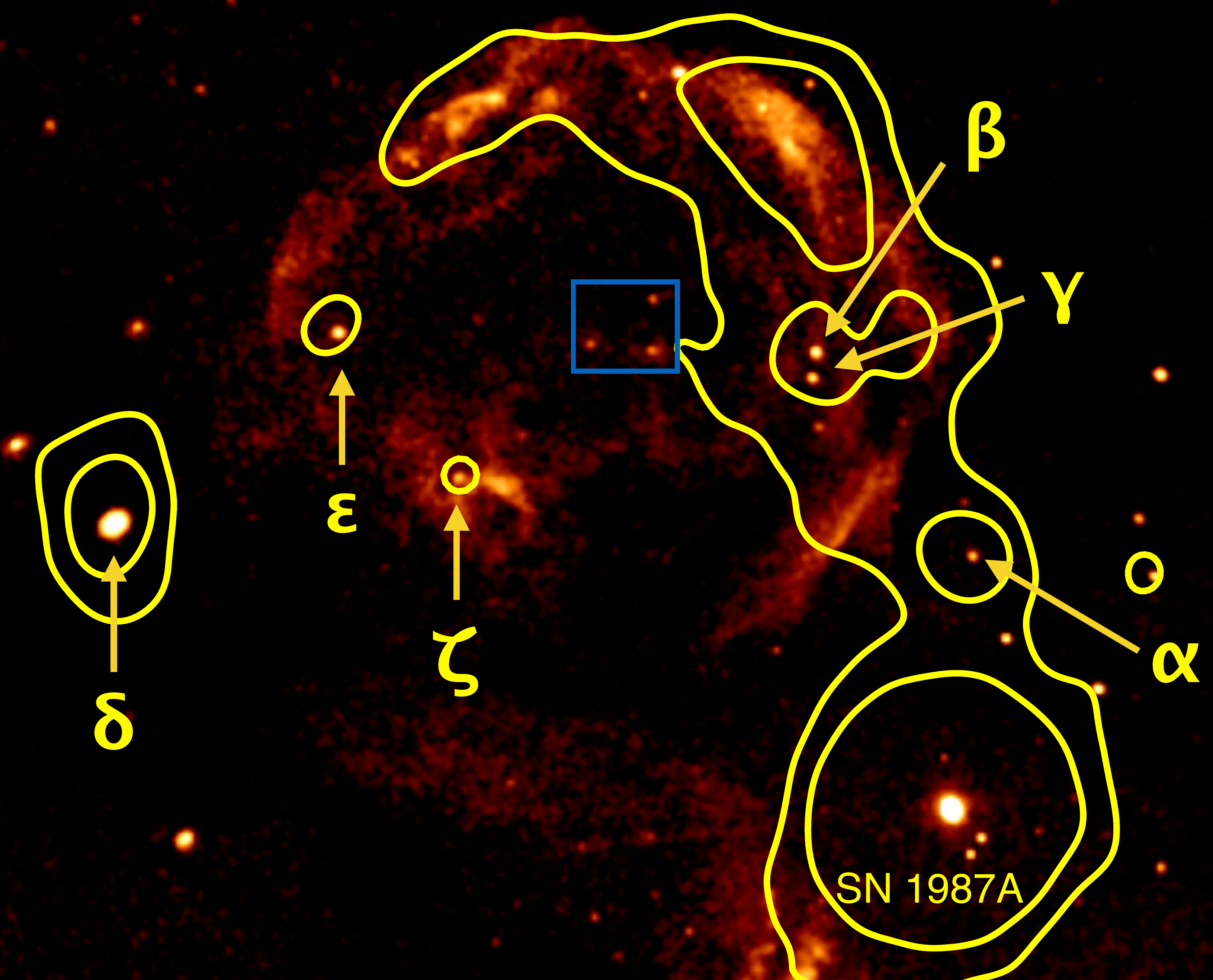}
\end{center}
\caption{Broad-band ($0.5-7.0$~keV), exposure-corrected {\it Chandra} image of 30~Dor~C using serendipitous observations from archival programs studying SN~1987A. The yellow contours are from the {\it NuSTAR} $3-20$ keV emission in Figure~\ref{fig:fullfield}. Multiple point sources resolved in the {\it Chandra} image coincide with emission detected by {\it NuSTAR}, including those labeled $\alpha$, $\beta$, $\gamma$, $\delta$, $\epsilon$, and $\zeta$ in this figure. Their identifications and properties in {\it Chandra} and {\it XMM-Newton} source catalogs are listed in Table~\ref{table:counterparts}. The three point sources enclosed by the blue box are associated with WR and massive-star clusters and are not detected by {\it NuSTAR}.}
\label{fig:chandra}
\end{figure}

\begin{deluxetable*}{lccccc}
\tablecolumns{6} \renewcommand{\arraystretch}{0.8}
\tablewidth{0pt} \tablecaption{Identified X-ray Point Sources Associated with {\it NuSTAR} Emission \label{table:counterparts}} 
\tablehead{\colhead{Source} & \colhead{CXO GSG\tablenotemark{a}} & \colhead{3XMM\tablenotemark{b}} & \colhead{$F_{\rm x}$\tablenotemark{c}} & \colhead{$L_{\rm x}$\tablenotemark{d}} & \colhead{Likely} \\
\colhead{} & \colhead{} & \colhead{} & \colhead{(erg cm$^{-2}$ s$^{-1}$)} & \colhead{(erg s$^{-1}$)} & \colhead{Associations}}
\startdata
$\alpha$ & J053525.7$-$691347 & J053525.9$-$691348 & 8.25$\times$10$^{-15}$ & 3.84$\times$10$^{33}$ & Unknown \\ 
$\beta$ & J053542.4$-$691152 & J053542.6$-$691153 & 3.17$\times$10$^{-14}$ & 9.72$\times$10$^{33}$ & Massive Star \\
$\gamma$ & J053542.9$-$691206 & -- & 1.03$\times$10$^{-14}$ & 5.10$\times$10$^{33}$ & AGN \\
$\delta$ & J053657.1$-$691328 & J053657.2$-$691329 & 4.79$\times$10$^{-13}$ & 1.74$\times$10$^{35}$ & X-ray Binary \\
$\epsilon$ & J053633.3$-$691140 & -- & 2.75$\times$10$^{-14}$ & 8.23$\times$10$^{33}$ & AGN \\
$\zeta$ & J053620.7$-$691303 & -- & 1.14$\times$10$^{-14}$ & 3.41$\times$10$^{33}$ & AGN or Stellar Remnant
\enddata
\tablenotetext{a}{Identifications from the {\it Chandra} ACIS GSG Point-Like X-ray Source Catalog \citep{wang16}. For the point sources $\epsilon$ and $\zeta$, the identifications are from \cite{bamba04}.}
\tablenotetext{b}{Identifications from the 3XMM DR6 version of the {\it XMM-Newton} Serendipitous Source Catalog \citep{rosen16}. Point sources $\gamma$ and $\epsilon$ do not have a coincident detection reported in any {\it XMM-Newton} catalogs.}
\tablenotetext{c}{Absorption-corrected X-ray flux in the 0.3--8.0~keV band from {\it Chandra}, as reported by \cite{wang16} for $\alpha$, $\beta$, $\gamma$, and $\delta$. They assumed Galactic absorption only and fit the spectrum with a power-law of index $\Gamma$ = 1.7. Absorption-corrected fluxes for sources $\epsilon$ and $\zeta$ have been computed for the 0.3--8.0~keV band using the spectral fit results reported in Table~2 of \cite{bamba04}.}
\tablenotetext{d}{X-ray luminosity assuming the object is in the LMC at a distance of $D =$~50~kpc \citep{pie19}. Note that Sources $\gamma$ and $\epsilon$ may be AGN, so their distances may be much larger and luminosities much greater.}
\end{deluxetable*}

In Figure~\ref{fig:threecolor}, we compare the $3-20$~keV background-subtracted {\it NuSTAR} image (in blue) with the 1.4~GHz radio morphology (in white contours; from \citealt{hughes07}) and the H$\alpha$ emission (in red; from \citealt{smith98}). The complete shell of 30~Dor~C is evident in the radio and had been reported at 843~MHz, 1.38~GHz, and 5.5~GHz frequencies previously by \cite{mills84}, \cite{kavanagh15}, and \cite{kavanagh18}, respectively. The radio coincidence with the hard X-rays is consistent with a synchrotron origin of the {\it NuSTAR}-detected emission. The H$\alpha$ image also shows a shell morphology, with a relatively narrow rim in the northwest, where 30~Dor~C is bright in hard X-rays. \cite{mathewson85} and \cite{kavanagh15} found that the radio spectral index is flatter in the western part of 30~Dor~C (with $-0.5 \lesssim \alpha \lesssim 0.5$), consistent with thermal radio emission and possibly due to contamination from a foreground molecular cloud. The eastern side has a steeper spectral index (with $-0.6 \gtrsim \alpha \gtrsim-2.2$) that these authors interpret as evidence of non-thermal radio emission there.

\subsubsection{Point Sources} 

Multiple point sources are also apparent in the {\it NuSTAR} images. To aid in the identification of counterparts, we compared the {\it NuSTAR} $3-20$ keV morphology to the broad-band ($0.5-7.0$~keV), exposure-corrected {\it Chandra} image of 30~Dor~C, as shown in Figure~\ref{fig:chandra}. Six locations (marked $\alpha$, $\beta$, $\gamma$, $\delta$, $\epsilon$, and $\zeta$ in Figure~\ref{fig:chandra}) of bright {\it NuSTAR} emission coincide with point sources that are evident in the {\it Chandra} image. Four of these point sources ($\alpha$, $\beta$, $\gamma$, and $\delta$) have been identified in X-ray catalogs \citep{evans10,rosen16,wang16}, and their names, fluxes, and luminosities (assuming the objects are in the LMC) are given in Table~\ref{table:counterparts}. Two additional {\it Chandra} point sources ($\epsilon$ and $\zeta$) are detected by {\it NuSTAR} that are not in {\it Chandra} or {\it XMM-Newton} catalogs, but \cite{bamba04} identified and characterized these objects in their study of 30~Dor~C (CXOU~J053633.3$-$691140 and CXOU~J3620.7$-$691303, respectively, in their Table~2). We include properties of these sources in Table~\ref{table:counterparts} as measured by \cite{bamba04}. Unfortunately, count statistics were not sufficient to model the {\it NuSTAR} spectra from these sources, and we discuss their likely associations based on {\it Chandra} and multiwavelength data below (as listed in Table~\ref{table:counterparts}).

Source~$\alpha$ has no known counterparts at other wavelengths besides the {\it Chandra} point source. The closest point source listed in the Simbad astronomical database \citep{wenger00} is a blue supergiant star (CPD--69 400) 9.4\arcsec\ away, much greater than the $\approx$0.5\arcsec\ point-spread function of {\it Chandra}, indicating that the soft X-rays originate from a different source. We note that Source~$\alpha$ is located at the center of an evacuated cavity in the H$\alpha$ image (see Figure~\ref{fig:threecolor}), suggesting that the hard X-ray emission may be associated with another massive star or diffuse gas that was shock-heated by stellar winds.

Source~$\beta$ was studied by \cite{bamba04} who extracted {\it Chandra} X-ray spectra and showed that the data were best-fit by a collisional equilibrium thermal plasma model (specifically an absorbed {\it mekal} component with abundances of 0.3 solar). Based on its location and spectral properties, \cite{bamba04} concluded that Source~$\beta$ is associated with Brey~58, a O3If$^{\ast}$/WN6 star \citep{massey00,neugent12}. 

Another point source, $\gamma$, is $\approx$15\arcsec\ south of Source~$\beta$. \cite{bamba04} stated that no optical or infrared counterpart is coincident with $\gamma$, and they suggested it is a background AGN or X-ray binary based on its hard spectra. 

\cite{lin12} classified Source~$\delta$ as a candidate compact-object binary based on its hardness ratio and its X-ray-to-infrared flux ratio in the 2XMMi-DR3 catalog (where the object is identified as 2XMM J053657.1$-$691328: \citealt{watson09}). 

Source~$\epsilon$ has no known counterparts at optical or infrared wavelengths, and it is not in any X-ray catalogs, although it is detected in both the {\it Chandra} and the {\it XMM-Newton} images (see Figure~\ref{fig:xmm_label}). \cite{bamba04} extracted {\it Chandra} spectra from Source $\epsilon$ and found the data to be best-fit by a power-law with a spectral index of $\Gamma$=1.8$^{+0.5}_{-0.3}$ and concluded it is most likely a background AGN.

Source~$\zeta$ also has no known counterparts at optical or infrared wavelengths, and it is located on the periphery of SNR~J0536$-$6913. \cite{bamba04} found Source~$\zeta$ to have a relatively hard spectrum, best-fit by a power-law with a spectral index $\Gamma = 1.9^{+0.5}_{-0.4}$. Based on these results, \cite{bamba04} suggest Source~$\zeta$ may be a background AGN or stellar remnant (e.g., a neutron star or black hole). The latter explanation is intriguing given Source~$\zeta$'s proximity to the young SNR. However, the velocity necessary for the object to travel from the SNR center to its current location\footnote{We assumed the position of Source~$\zeta$ as RA = 05$^{\rm h}$36$^{\rm m}$20.477$^{\rm s}$, Dec=$-$69$^{\rm d}$13$^{\rm m}$02.70$^{\rm s}$, and we employed the SNR center of RA=05$^{\rm h}$36$^{\rm m}$17.0$^{\rm s}$, Dec=$-$69$^{\rm d}$13$^{\rm m}$28.0$^{\rm s}$ reported by \cite{kavanagh15}.} would be 1500--3400~km~s$^{-1}$ (for ages of 2.2--4.9~kyr), greater than velocities reported for the known pulsar population \citep{faucher06}.

Three central point sources evident in the {\it Chandra} image (in the blue box in Figure~\ref{fig:chandra}) are not clearly detected in the {\it NuSTAR} observations (though some diffuse emission is detected in the west of this region). These sources (going clockwise) are associated with the WR star Brey~57 and massive star clusters identified by \cite{lortet84} as $\gamma$ and $\beta$. The latter two sources were detected and analyzed by \citealt{bamba04} (their sources \#4 and 3, respectively) and found to have X-ray luminosities (in the 0.5--9~keV band) of $\approx(1-2)\times10^{33}$~erg~s$^{-1}$. Thus, these sources are fainter than those detected with {\it NuSTAR} and listed in Table~\ref{table:counterparts}.

\subsection{Spectroscopy} \label{sec:spectra}

\begin{figure}
\begin{center}
\includegraphics[width=\columnwidth]{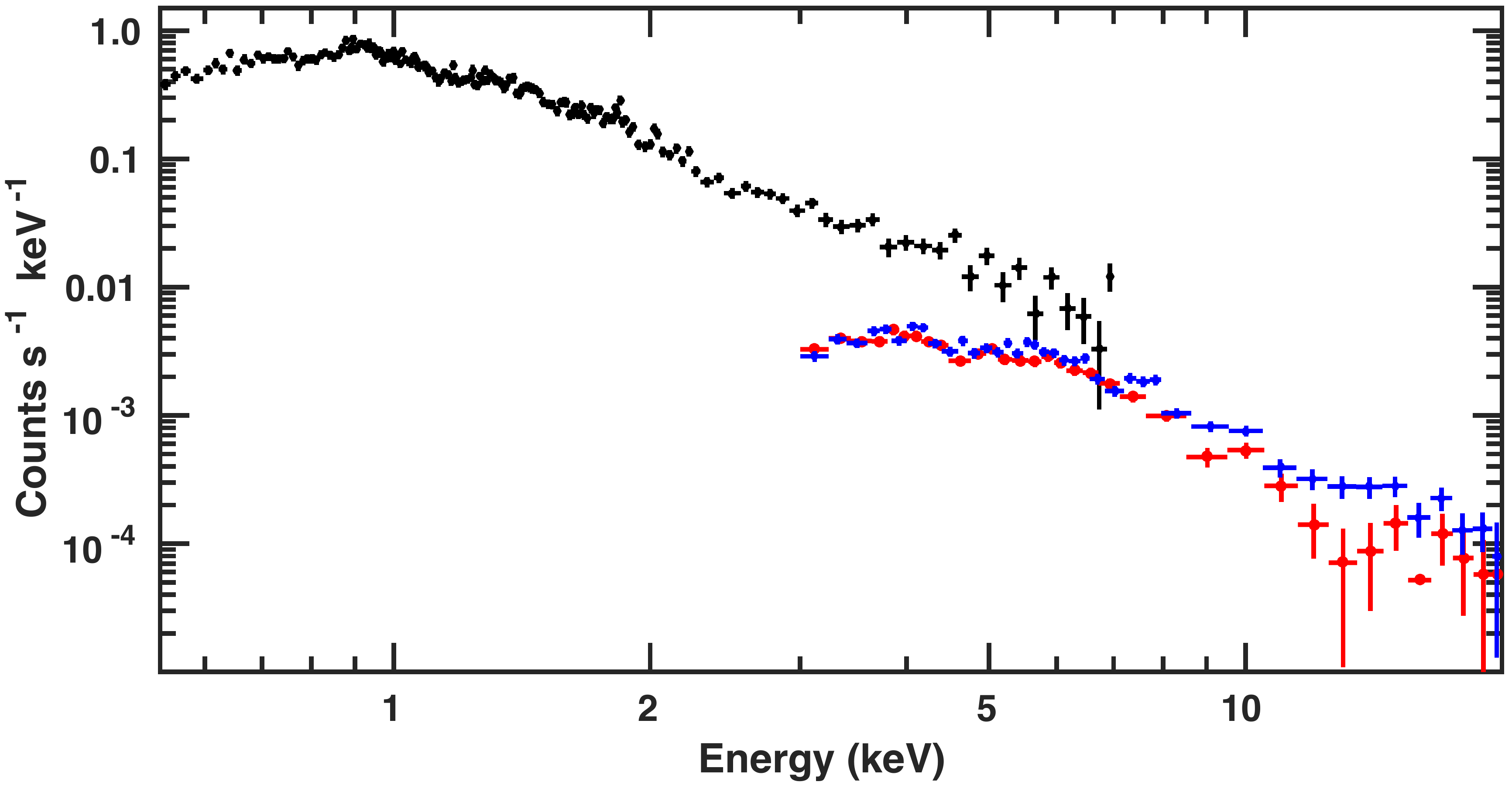}
\end{center}
\caption{Background-subtracted {\it XMM-Newton} EPIC-pn and {\it NuSTAR} X-ray spectra from the entirety of 30~Dor~C. {\it NuSTAR} spectra were extracted from the observations that have 30~Dor~C completely imaged (observations labeled \#1--5, \#14, and \#16--17 in Figure~\ref{fig:expmaps}), and data from FPMA (black) and FPMB (dark blue) were combined to produce one {\it NuSTAR} spectrum for each module. As discussed in Section~\ref{sec:nustardata}, the FPMB data had substantial stray-light contamination, limiting the locations where background spectra could be extracted. Thus, the differences between the FPMA and FPMB spectra likely arise from the challenges of extracting and modeling the background in the FPMB data.}
\label{fig:globalspectrum}
\end{figure}

Figure~\ref{fig:globalspectrum} shows the background-subtracted {\it XMM-Newton} EPIC-pn and {\it NuSTAR} spectra from the entirety of 30~Dor~C (including point sources and MCSNR~J0536$-$6913). Spectra were extracted from the {\it NuSTAR} pointings that fully imaged 30~Dor~C (those labeled \#1--5, \#14, \#16--17 in Figure~\ref{fig:expmaps}), and the data from each FPMA and FPMB observation were combined to produce one {\it NuSTAR} spectrum for each module. 30~Dor~C was detected up to $\sim$20~keV, and the background dominates $\gtrsim$20~keV. 

\begin{figure}[!t]
\begin{center}
\includegraphics[width=\columnwidth]{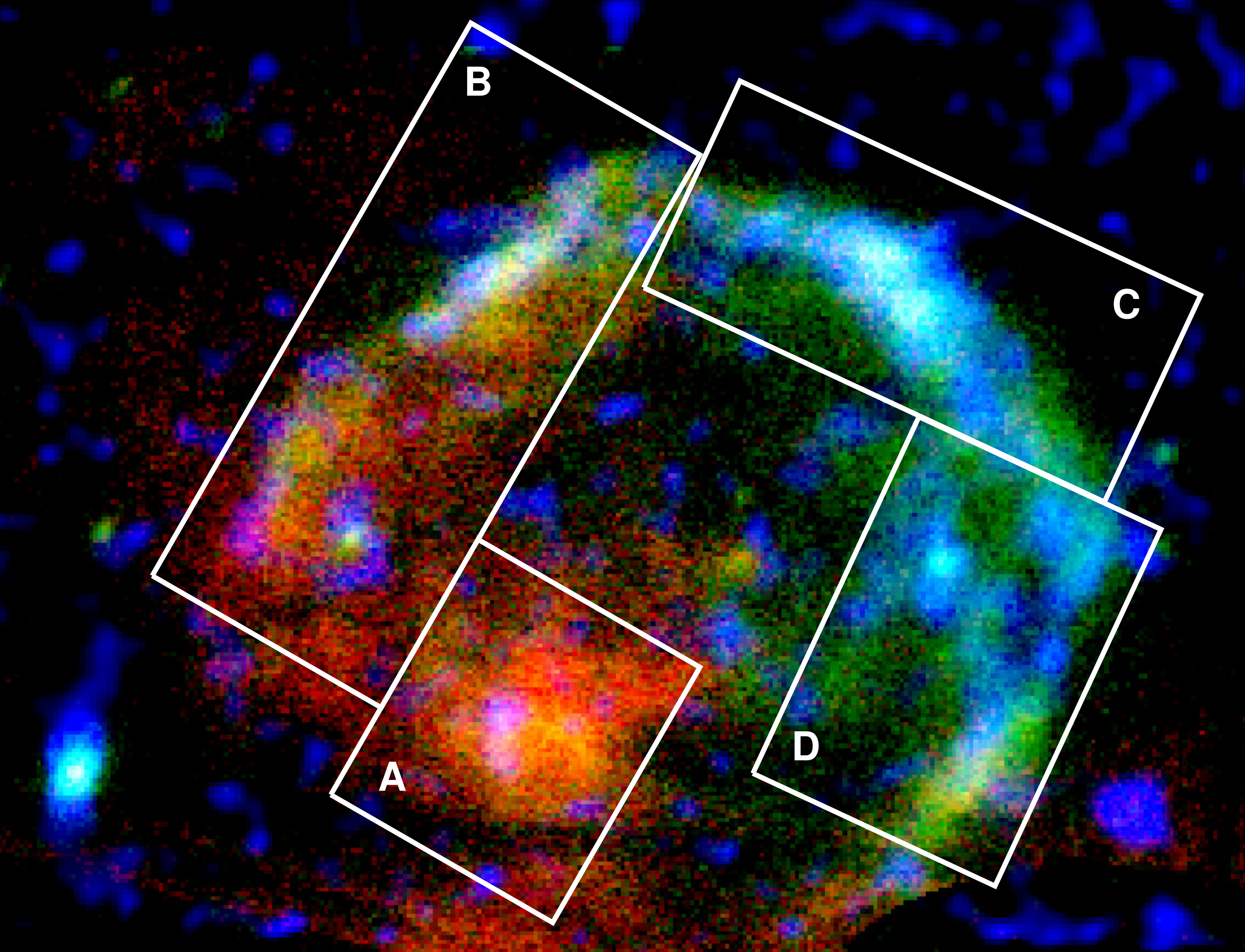}
\end{center} 
\caption{Three-color image produced using {\it XMM-Newton} and {\it NuSTAR} data of 30~Dor~C, with regions of spectral extraction labeled. The image has {\it XMM-Newton} data in red (0.3--1.0 keV) and in green (1--3 keV) and {\it NuSTAR} data in blue (3--20 keV). Spectra were extracted from four regions (labeled A, B, C, and D) of 30~Dor~C. These regions roughly correspond to those analyzed by \cite{kavanagh15} and \cite{bamba04} using {\it XMM-Newton} and {\it Chandra} data, respectively.}
 \label{fig:xmm_label}
\end{figure}

We also extracted {\it XMM-Newton} and {\it NuSTAR} spectra from three regions of 30~Dor~C's shell as well as one region enclosing the MCSNR J0536$-$6913; Figure~\ref{fig:xmm_label} denotes these locations. In order to facilitate comparison to previous X-ray studies of 30~Dor~C (specifically, \citealt{bamba04,yamaguchi09,kavanagh15}), we selected similar regions as those works. 

We do not detect X-rays above 8~keV from MCSNR J0536$-$6913 with {\it NuSTAR}. Thus, we did not analyze the spectra from Region~A further as the region has already been investigated with {\it Chandra} and {\it XMM-Newton}, and we cannot set any additional constraints with the {\it NuSTAR} data. \cite{kavanagh15} fit the {\it XMM-Newton} spectra from the SNR and showed that it was best described by a model with two thermal plasmas representing the contributions of ejecta and shock-heated ISM. They found enhanced abundances of intermediate-mass elements that suggested a core-collapse origin, and they estimated an age of $\sim$2.2--4.9~kyr for the SNR. 

For the three other regions, we fit the {\it XMM-Newton} and {\it NuSTAR} spectra jointly over the 0.5--8.0~keV and 3--20~keV range, respectively, with an absorbed {\it srcut} model and adding an optically-thin thermal plasma ({\it apec}) component to test if it improved the fits. As discussed in Section~\ref{sec:xmmdata}, in the case of the {\it XMM-Newton} data, we modeled the background using the models of \cite{maggi16}. For the source {\it apec} components, we froze the abundance to 0.5$Z_{\sun}$, representative of the LMC ISM \citep{russell92}\footnote{We note that adopting a {\it vapec} component with the individual elemental abundances of \cite{maggi16} yielded the same fit results as the {\it apec} models.}. We fit the data from all instruments jointly by including a multiplicative factor (with the XSPEC component {\it const}) for each dataset that is allowed to vary while all other model parameters were required to be the same. We had two source absorption components: one to account for the Galactic absorption $N_{\rm H}$ (with the XSPEC model {\it phabs}, assuming cross-sections from \citealt{verner96} and the solar abundances of \citealt{wilm00} to be consistent with \citealt{kavanagh15}), and another for the LMC's intrinsic absorption $N_{\rm H, LMC}$ (with the XSPEC model {\it vphabs} and adopting the LMC ISM abundances of \citealt{maggi16})\footnote{We also performed the fits using the XSPEC absorption model {\it tbabs}, and the fit results were the same as with {\it phabs}.}. 

The {\it srcut} model describes the source spectrum as the synchrotron emission from a power-law energy distribution of electrons with an exponential cutoff at energy $E_{\rm max}$ \citep{reynolds99}. This model falls off considerably more slowly than an exponential (roughly as $\exp(-\sqrt{\nu/\nu_{\rm rolloff}})$) where the rolloff frequency $\nu_{\rm rolloff}$ is the characteristic or critical synchrotron frequency of electrons with energy $E_{\rm max}$ ($\nu_{\rm c}$ in the notation of \citealt{pach70}). The rolloff photon energy $E_{\rm rolloff} = h \nu_{\rm rolloff}$ is thus related to $E_{\rm max}$ by 

\begin{equation}
E_{\rm max} = 120  \bigg(\frac{h \nu_{\rm rolloff}}{{\rm 1~keV}} \bigg)^{1/2} \bigg(\frac{B}{\mu{\rm G}} \bigg)^{-1/2}~~{\rm TeV}
\label{eq:Emax}
\end{equation}

\noindent
where $B$ is the magnetic field strength. In XSPEC, the {\it srcut} model has three parameters: the rolloff frequency $\nu_{\rm rolloff}$, the mean radio-to-X-ray spectral index $\alpha$, and the 1~GHz radio flux density $F_{\rm 1GHz}$. To limit the free parameters in the fit, we estimated $F_{\rm 1GHz}$ of each region by measuring the 1.4~GHz flux density at those locations in the \cite{hughes07} survey of the LMC with the Australia Telescope Compact Array (ATCA) and the Parkes Telescope (see the white contours in Figure~\ref{fig:threecolor}). We assume a radio spectral index of $\alpha=-0.65$ to derive $F_{\rm 1GHz}$ of each region, and the values are listed in Table~\ref{table:spectralresults}. The radio spectral index of $\alpha=-0.65$ is an intermediate value between the radio spectral index of $\alpha=-0.5$ adopted by \cite{bamba04} and the best-fit spectral index of $\alpha=-0.75$ found by \cite{kavanagh15} in the northeastern part of 30~Dor~C. However, the radio spectrum is actually concave, and $\alpha$ may flatten at higher frequencies \citep{reynolds92}. We have likely overestimated the flux density at 1~GHz arising from synchrotron radiation given that free-free emission also contributes at these wavelengths. 

\begin{deluxetable*}{lcccccrcc}
\tablecolumns{9}
\setlength{\tabcolsep}{0.0in} \renewcommand{\arraystretch}{1.0}
\tablewidth{0pt} \tablecaption{Spectral Results with a SRCUT Component\tablenotemark{a} \label{table:spectralresults}} 
\tablehead{\colhead{Region} & \colhead{$N_{\rm H, LMC}$} & \colhead{$kT$} & \colhead{norm\tablenotemark{b}} & \colhead{$\nu_{\rm rolloff}$} & \colhead{$F_{\rm 1~GHz}$} & \colhead{$\chi^{2}$/d.o.f.} & \colhead{$F_{\rm X}$\tablenotemark{c}} & \colhead{$F_{\rm nt}$\tablenotemark{d}} \\
\colhead{} & \colhead{($\times10^{21}$ cm$^{-2}$)} & \colhead{(keV)} & \colhead{($\times10^{-4}$)} & \colhead{($\times10^{17}$~Hz)} & \colhead{(Jy)} & \colhead{} & \colhead{($\times10^{-12}$ erg cm$^{-2}$ s$^{-1}$)} & \colhead{($\times10^{-12}$ erg cm$^{-2}$ s$^{-1}$)}}
\startdata
B & 0.9$\pm$0.1 & 0.86$\pm$0.01 & 8.8$^{+0.5}_{-1.3}$ & 4.0$^{+0.3}_{-0.8}$ & 0.66 & 12842/9643 & 1.9$^{+0.6}_{-0.7}$ & 1.5$\pm$0.5 \\
C & 8.0$^{+0.4}_{-0.3}$ & -- & -- &  3.2$^{+0.2}_{-0.3}$ & 0.43 & 9971/9297 & 1.9$\pm$0.3 & 1.9$\pm$0.3 \\
D & 2.4$\pm$0.1 & 0.86$^{+0.02}_{-0.01}$ & 7.2$^{+0.3}_{-0.4}$ & 8.4$^{+0.4}_{-0.6}$ & 0.72 & 12250/10112 & 1.8$\pm$0.4 & 1.6$\pm$0.4 \\    
%B & 2.4$\pm$0.1 & 0.86$\pm$0.01 & 6.7$\pm$0.5 & 3.3$^{+0.7}_{-0.3}$ & 0.66 & 10622/9679 & 1.6$\pm$0.4 & 1.4$\pm$0.4 \\
%C & 12.2$^{+0.3}_{-0.4}$ & -- & -- & 2.2$^{+0.3}_{-0.1}$ & 0.43 & 9144/9332 &  2.1$\pm$0.2 & 2.1$\pm$0.2 \\
%D & 4.3$\pm$0.1 & 0.84$\pm$0.02 & 6.6$\pm$0.5  & 6.4$^{+0.5}_{-0.4}$ & 0.72 & 12083/10864 & 2.2$\pm$0.6 & 1.6$\pm$0.2 
\enddata
\tablenotetext{a}{Error bars represent the 90\% confidence range.}
\tablenotetext{b}{Normalization of the {\it apec} component, defined as norm $= \frac{10^{-14}}{4 \pi D^{2}} \int n_{\rm e } n_{\rm H} dV$, where $D$ is the distance to the source in cm, $n_{\rm e}$ and $n_{\rm H}$ are the electron and hydrogen densities in cm$^{-3}$, respectively, and $V$ is the volume.}
\tablenotetext{c}{Total unabsorbed flux in the 0.5--20~keV band.}
\tablenotetext{d}{Unabsorbed flux from the {\it srcut} component in the 0.5--20~keV band.}
\end{deluxetable*}

\begin{deluxetable*}{lcccccrcc}
\tablecolumns{9}
\setlength{\tabcolsep}{0.0in} \renewcommand{\arraystretch}{1.0}
\tablewidth{0pt} \tablecaption{Spectral Results with a Power-Law Component\tablenotemark{a} \label{table:spectralresults2}} 
\tablehead{\colhead{Region} & \colhead{$N_{\rm H, LMC}$} & \colhead{$kT$} & \colhead{norm\tablenotemark{b}} & \colhead{$\Gamma$} & \colhead{norm$_{\rm PL}$\tablenotemark{c}} & \colhead{$\chi^{2}$/d.o.f.} & \colhead{$F_{\rm X}$\tablenotemark{d}} & \colhead{$F_{\rm nt}$\tablenotemark{e}} \\
\colhead{} & \colhead{($\times10^{21}$ cm$^{-2}$)} & \colhead{(keV)} & \colhead{($\times10^{-4}$)} & \colhead{} & \colhead{($\times10^{-4}$)} & \colhead{} & \colhead{($\times10^{-12}$ erg cm$^{-2}$ s$^{-1}$)} & \colhead{($\times10^{-12}$ erg cm$^{-2}$ s$^{-1}$)}} 
\startdata
B & 1.4$\pm$0.1 & 0.86$\pm$0.01 & 0.2\tablenotemark{f} & 2.26$\pm$0.03 & 2.3\tablenotemark{f} & 12819/9701 & 2.0$\pm$0.7 & 1.6$\pm$0.6 \\
C & 10.3$\pm$0.5 & -- & -- & 2.39$\pm$0.03 & 6.3\tablenotemark{f} & 10036/9295 & 2.2$\pm$0.2 & 2.2$\pm$0.2 \\
D & 2.8$\pm$0.1 & 0.86$\pm$0.01 & 9.2\tablenotemark{f} & 2.12$\pm$0.02 & 25\tablenotemark{f} & 12410/10110 & 1.9$\pm$0.3 & 1.7$^{+0.3}_{-0.2}$ \\
%B & 3.0$\pm$0.2 & 0.86$\pm$0.01 & 6.6\tablenotemark{f} & 2.30$^{+0.04}_{-0.03}$ & 1.2\tablenotemark{f} & 10650/9676 & 1.9$\pm$0.2 & 1.6$\pm$0.2  \\ %r3
%C & 14.2$\pm$0.4 & -- & -- & 2.49$\pm$0.03 & 6.6\tablenotemark{f} & 9144/9331 & 2.5$\pm$0.2 & 2.5$\pm$0.2 \\
%D & 4.9$\pm$0.2 & 0.83$\pm$0.02 & 5.4\tablenotemark{f} & 2.20$\pm$0.02 & 14.8\tablenotemark{f} & 12224/10863 & 2.0$\pm$0.3 & 1.6$\pm$0.2
\enddata
\tablenotetext{a}{Error bars represent the 90\% confidence range.}
\tablenotetext{b}{Normalization of the {\it apec} component, defined as norm $= \frac{10^{-14}}{4 \pi D^{2}} \int n_{\rm e } n_{\rm H} dV$, where $D$ is the distance to the source in cm, $n_{\rm e}$ and $n_{\rm H}$ are the electron and hydrogen densities in cm$^{-3}$, respectively, and $V$ is the volume.}
\tablenotetext{c}{Normalization of the power-law component, in units of photons keV$^{-1}$ cm$^{-2}$ s$^{-1}$ at 1~keV.}
\tablenotetext{d}{Total unabsorbed flux in the 0.5--20 keV band.}
\tablenotetext{e}{Unabsorbed flux from the power-law component in the 0.5--20~keV band.}
\tablenotetext{f}{Denotes parameters that are not constrained in the model, so no 90\% confidence range is given.}
\end{deluxetable*}

\begin{figure}
\begin{center}
\includegraphics[width=0.35\textwidth]{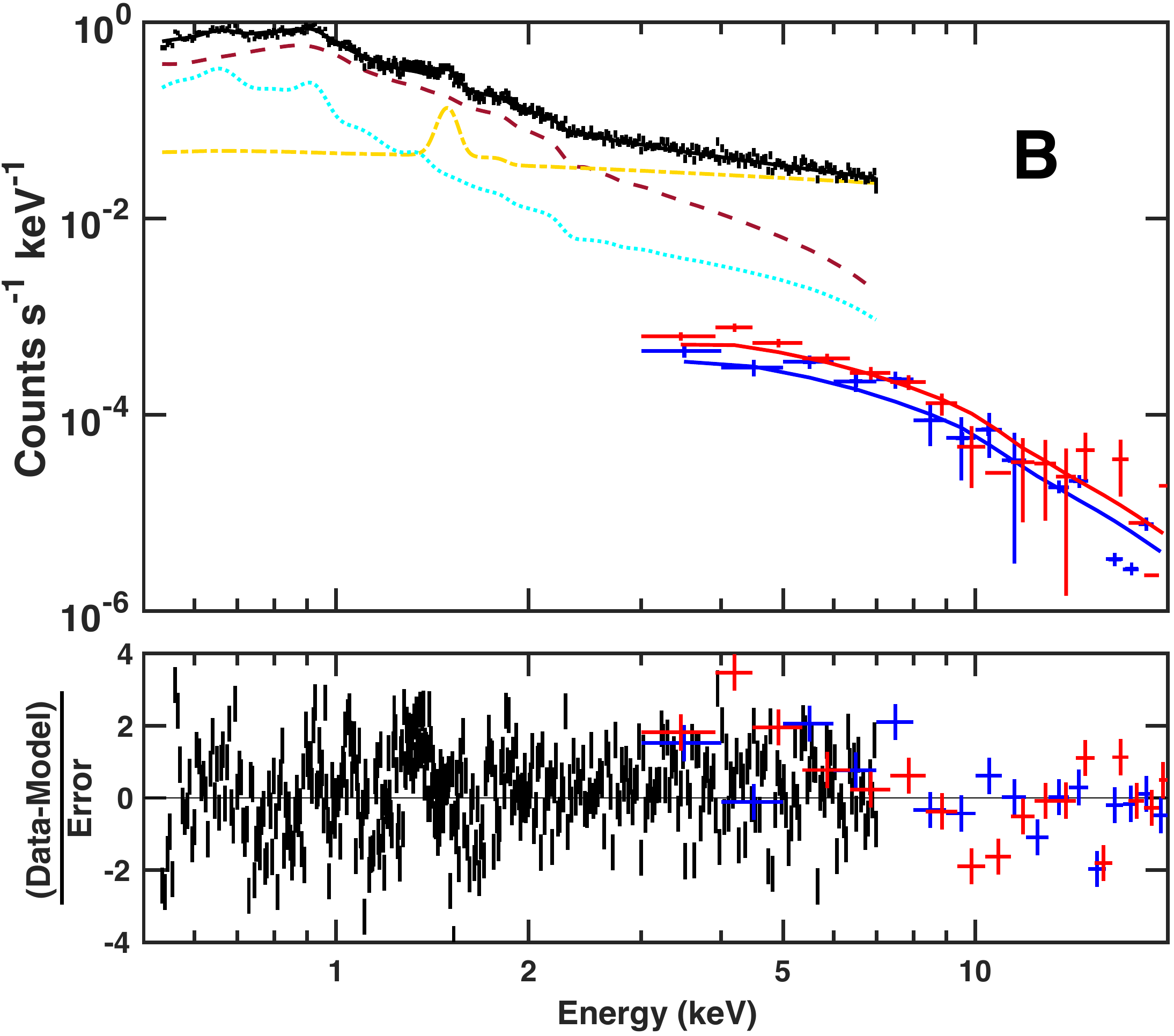}
\includegraphics[width=0.35\textwidth]{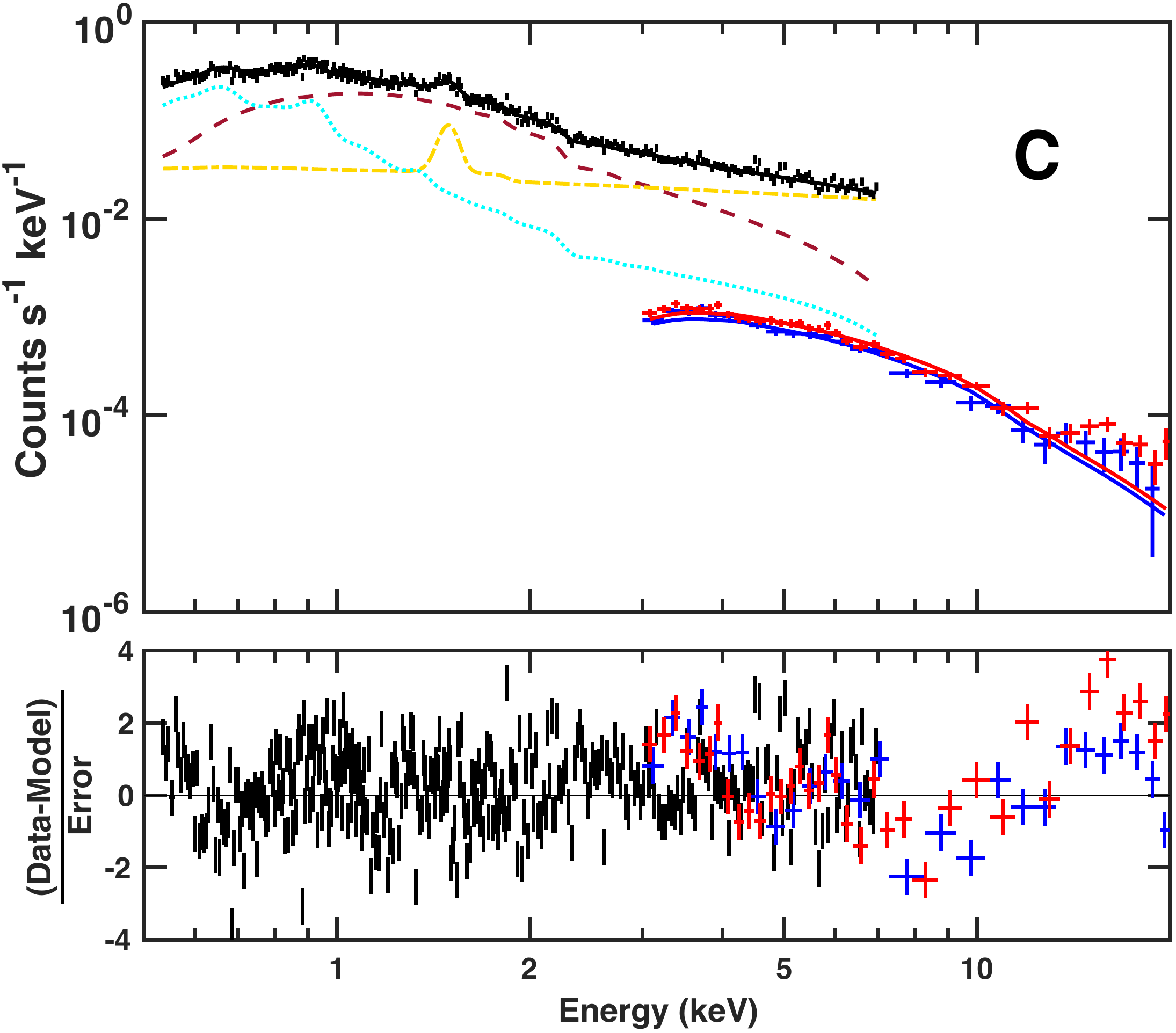}
\includegraphics[width=0.35\textwidth]{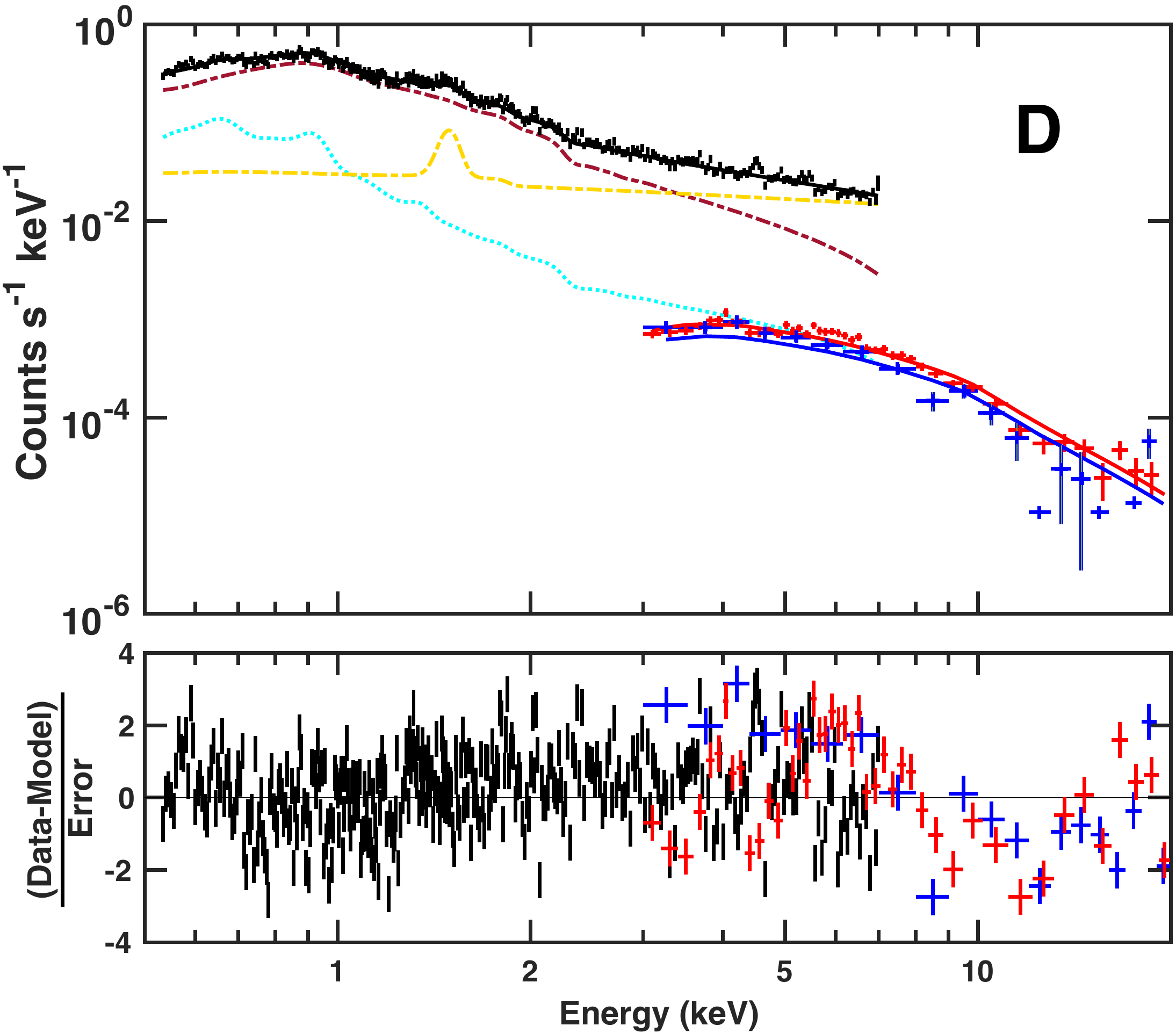}
\end{center}
\caption{{\it XMM-Newton} and {\it NuSTAR} X-ray spectra from Regions B (top), C (middle), and D (bottom), as labeled in Figure~\ref{fig:xmm_label}. For simplicity/clarity, we plot only one {\it XMM-Newton} observation (the EPIC-pn data from ObsID 0506220101, which had the longest effective exposure of $\approx$97~ks) in black. The black solid line represents the total {\it XMM-Newton} model, where the burgundy dashed line is the source emission, the cyan dotted line is the AXB component, and the yellow dash-dotted line is the instrumental background component. The combined FPMA data are in blue, and the FPMB data are in red. The source spectra of Region~C were fit with an absorbed {\it srcut} model (solid line), while those of Regions~B and ~D were best-fit by an absorbed {\it srcut} component plus a thermal {\it apec} component.}
 \label{fig:spectra}
\end{figure}

Figure~\ref{fig:spectra} shows the spectra and best fits from each of the three 30~Dor~C shell regions. Table~\ref{table:spectralresults} lists the results, including $N_{\rm H, LMC}$, $\nu_{\rm rolloff}$, $\chi^{2}$ and the degrees of freedom (d.o.f.). Regions B, C, and D were inadequately fit by a single thermal component, producing $\chi^{2}$/d.o.f. $\gtrsim$3 in all cases. Region~C was best fit with a single {\it srcut} component, with $\nu_{\rm rolloff} = (3.2^{+0.2}_{-0.3})\times10^{17}$~Hz and $\chi^{2}$/d.o.f. = 9971/9297. We fit Regions~B and~D initially with a single {\it srcut} component, and we noted large residuals below $\sim$1~keV, though the fits were statistically acceptable, with $\chi^{2}$/d.o.f.=12842/9643 for Region~B and $\chi^{2}$/d.o.f.=12250/10112 for Region~D. The addition of an {\it apec} component to these models improved the fits, as listed in Table~\ref{table:spectralresults}. If we instead adopted $\alpha=-0.5$ or $\alpha=-0.75$ in our analysis, the best-fit values of $\nu_{\rm rolloff}$ decreased or increased by $\sim$60\%, respectively. However, fits with $\alpha = -0.5$ and $\alpha = -0.75$ yielded the same relative contribution of the non-thermal to thermal components in the 0.5--20~keV band in Regions~B and D  (82\% and 90\%, respectively).

Past X-ray studies of 30~Dor~C have found differing results on the relative contribution of the thermal and non-thermal emission across the shell. Generally, previous investigations \citep{bamba04,yamaguchi09,kavanagh15,babazaki18} agree that Region~C only requires a single non-thermal component to adequately fit the X-ray spectra there. Additionally, all of these works except \cite{bamba04} found that Region~B necessitates a thermal plasma component (in addition to the non-thermal component) to account for the residuals at soft X-ray energies. However, the estimates of the temperature of that thermal plasma in Region~B differ: e.g., $\sim$0.2 keV \citep{babazaki18}, $\sim$0.3~keV \citep{kavanagh15}, $\sim$0.7~keV \citep{yamaguchi09}, and 0.86$\pm$0.01~keV in this work. Finally, our result that Region~D requires a thermal component has not been found in prior studies that analyzed the X-ray spectra there \citep{bamba04,kavanagh15,babazaki18}. The disparate results likely arise from authors using slightly different spectral extraction regions as well as different XSPEC components to describe the non-thermal (i.e., {\it srcut} versus {\it powerlaw}) and the thermal (e.g., {\it apec} versus {\it vapec}, with variable abundances for intermediate-mass elements) contributions.

We also attempted to fit the source spectra with a power-law rather than the {\it srcut} component to account for the non-thermal emission. The results are listed in Table~\ref{table:spectralresults2}. In all regions, the power-law model was as successful as the {\it srcut} model in fitting the data. In past X-ray work on 30~Dor~C, \cite{yamaguchi09} statistically favored the {\it srcut} models over a simple power-law in our Region~C, whereas \cite{bamba04} and \cite{kavanagh15} reported that they could not distinguish between these two models statistically. 

\section{Discussion} \label{sec:discuss}

We have shown that the complete shell of 30~Dor~C emits hard, non-thermal X-rays. We fit {\it XMM-Newton} and {\it NuSTAR} spectra at three locations around the shell, and a power-law or a {\it srcut} component was comparably successful at modeling these data. Given the best-fit $\nu_{\rm rolloff}\sim(3-8)\times10^{17}$~Hz of the {\it srcut} components, we can estimate the maximum energy of the accelerated electrons $E_{\rm max}$ using Equation~\ref{eq:Emax}. Assuming the total B-field strength of $B \approx4~\mu$G\footnote{This value is consistent with the values of $B = 3-20~\mu$G found by \cite{kavanagh18} as well as the ISM LMC B-field strength of $B = 1~\mu$G from \cite{gaensler05} and assuming a compression ratio of 4.}, we find \hbox{$E_{\rm max} \approx 70-110$}~TeV. If we instead adopt the 90\% confidence limit of $B \lesssim$ 40~$\mu$G found by \cite{kavanagh18}, we get $E_{\rm max} \approx 20-35$~TeV. These $E_{\rm max}$ values are comparable to those found in young SNRs (e.g., \citealt{reynolds99,lopez15}). 

Our estimates of $E_{\rm max}$ are also roughly consistent with the TeV detection of 30~Dor~C with HESS up to $\sim$20~TeV \citep{HESS15}. The TeV spectrum was best-fit by a power-law with photon index of $\Gamma =2.6\pm0.2$, yielding a total $L_{\gamma} = (0.9\pm0.2)\times10^{35}$~erg~s$^{-1}$ in the $1-10$~TeV band. Although it is unclear whether the emission was produced by a hadronic or leptonic cosmic-ray population (\citealt{kavanagh18} argue for a leptonic origin based on the $B$-field strength), the signal was localized to a central/northwest region of 30~Dor~C where the massive stars associated with LH~90 are located (see Figure~1 of \citealt{HESS15}), including part of our Region~C. 

In the context of SNRs, relativistic electrons are thought to be accelerated by diffusive shock acceleration (DSA; \citealt{bell04}), and shocks of velocities $\gtrsim$1,000~km~s$^{-1}$ are necessary for particles to reach TeV energies. In the case of 30~Dor~C, the expansion velocity of the H$\alpha$ shell is only $\lesssim$100~km~s$^{-1}$ \citep{dunne01,kavanagh18}, though the H$\alpha$ shell velocity does not reflect the shock responsible for the synchrotron emission. \cite{kavanagh18} suggested that the shocks are stalled in regions where it encountered the H$\alpha$ shell, and it expands at faster velocities through gaps in the shell elsewhere. This explanation is consistent with the anti-correlation of the H$\alpha$ and X-ray they found in the northeast and northwest (and that is evident when comparing the H$\alpha$ with the {\it NuSTAR} emission in Figure~\ref{fig:threecolor}).

Generally, particle acceleration in 30~Dor~C and other SBs is distinct from DSA in a single, isolated SNR. For example, \cite{par04} reviewed the collective effects of particle acceleration following numerous SNe and highlighted the differences from the isolated SN case. They showed that massive stars in OB associations are close enough that their winds interact, generating strong turbulence and magnetohydrodynamic (MHD) waves. They demonstrated that several mechanisms (e.g., wind-wind interactions, shock-cloud interactions, shock distortions) maintain turbulence and magnetic inhomogeneities in the SB interior, facilitating efficient turbulent acceleration. Finally, they found that SBs can yield particles with energies up to $\sim$10$^{17}$~eV through repeated acceleration. \cite{bykov01} showed that \hbox{$\sim10-30$\%} of a SB's turbulent energy can be transferred to low-energy, non-thermal particles to accelerate them, and \cite{butt08} suggested that up to one-third of the energy injected by stellar winds and SNe can go into accelerating cosmic rays. 

\cite{ferrand10} used semi-analytical models of particle acceleration (both DSA and stochastic reacceleration) inside SBs to investigate the shape of the resulting cosmic-ray spectra and their temporal evolution. The spectra depend on star cluster and SB parameters (e.g., number of stars $N_{\star}$, the external scale of turbulence $\lambda_{\rm max}$, the size of the acceleration region $x_{\rm acc}$, and the B-field strength), and \cite{ferrand10} quantified the spectral hardness using the dimensionless parameter $\theta^{\star}$, which is roughly the ratio of the stochastic reacceleration time to the escape time. Low $\theta^{\star}$ implies that reacceleration is faster than escape, leading to hard spectra, whereas high $\theta^{\star}$ means that escape is faster than reacceleration, resulting in softer spectra where particles escape quickly after being accelerated by SN shocks.

{\cite{ferrand10} estimated $\theta^{\star}$ for a sample of Milky Way and LMC star clusters and found a range, with $\theta^{\star} = $ 10$^{2}$--10$^{6}$ assuming $B = 1~\mu$G and a turbulence index of $q = 5/3$ or $\theta^{\star}$ = 10$^{-3}$--10$^{0}$ for $B = 10~\mu$G and a turbulence index of $q = 3/2$. For comparison, we find that LH~90 (the central star cluster of 30~Dor~C) has $\theta^{\star}$ = 10$^{4}$ or $\theta^{\star}$ = 10$^{-1}$ in these two cases, respectively, where we have adopted $N_{\star} = 33$ \citep{testor93}, $\lambda_{\rm max} = 9$~pc (the separation of the 33 stars in LH~90 given the star cluster radius of 29~pc), $x_{\rm acc} = 47.5$~pc (the radius of the 30~Dor~C shell; \citealt{dunne01}), and an ambient density of $n =  10^{-2}$~cm$^{-3}$. Thus, it appears that 30~Dor~C is generally consistent with the other sources considered by \cite{ferrand10}.

We consider whether the energy injection from the stellar population and SNe in 30~Dor~C can account for the observed non-thermal luminosity of $L_{\rm nt} \approx (1.5\pm0.2)\times10^{36}$~erg~s$^{-1}$ (the sum of Regions~B, C, and D) following a similar calculation from \cite{kavanagh15}. LH~90, the star cluster powering 30~Dor~C, has 26 O-stars and 7 WR stars \citep{testor93}. \cite{smith04} estimated the wind luminosity from the O-stars is $(1-7)\times10^{37}$ erg s$^{-1}$, and \cite{kavanagh15} calculated a luminosity of $\sim$5$\times10^{38}$ erg s$^{-1}$ from the 7 WR stars. Assuming the WR lifetimes are $\sim7\times10^{5}$~years \citep{leitherer97} and averaging over the age of the superbubble ($\sim$4~Myr), the combined luminosity from the O- and WR stars over a 4-Myr timescale is $(1-1.6)\times10^{38}$~erg~s$^{-1}$. If $5-6$ SNe have also occurred in the region \citep{smith04}, then SNe have contributed $(4-5)\times10^{37}$~erg~s$^{-1}$ as well, adopting the standard 10$^{51}$~erg of kinetic energy per explosion. Thus, the total energy input from the stellar population and SNe is $(1.4-2.1)\times10^{37}$~erg~s$^{-1}$. Assuming an efficiency of 15\% for the transfer of the SB's turbulent energy to non-thermal particles (from Figure~1 of \citealt{bykov01} for an age of 4~Myr), $\approx(2-3)\times10^{37}$~erg~s$^{-1}$ would be available to power synchrotron emission. Given that this value is an order of magnitude above $L_{\rm nt}$, it seems that the non-thermal particles have sufficient energy to account for the observed flux.

\section{Conclusions} \label{sec:conclusions}

We have presented evidence of particle acceleration in the superbubble 30~Dor~C using hard X-ray images and spectra from targeted and serendipitous {\it NuSTAR} and {\it XMM-Newton} observations. The complete shell of the SB is detected up to $\sim$20~keV, and the young SNR MCSNR~J0536$-$6913 is detected up to 8~keV. Additionally, hard X-ray emission is evident at locations of six point sources previously identified with {\it Chandra} and {\it XMM-Newton}, and we discussed the possible associations of these objects with massive star clusters, AGN, and stellar remnants. We extracted {\it NuSTAR} and {\it XMM-Newton} spectra at three locations around the 30~Dor~C shell and modeled them using a non-thermal ({\it srcut} or power-law) component, adding a thermal ({\it apec} or {\it vpshock}) component as needed. All three regions have predominantly non-thermal emission, and two regions (the east and west side of the SB) have some thermal emission as well. From the {\it srcut} models, we find best-fit rolloff frequencies of $\nu_{\rm rolloff} \sim (3-8)\times10^{17}$~Hz, which correspond to maximum electron energies of $E_{\rm max} \approx 70-110$~TeV. 

In addition to diffusive shock acceleration from individual SNR shocks, superbubbles may re-accelerate low-energy particles via turbulence and MHD waves, transferring tens of percent of the SB's turbulent energy to those particles. We show that the mechanical energy from the stellar population and previous SN explosions in the bubble's interior is sufficient to account for the observed non-thermal flux.

\acknowledgements

We thank Parviz Ghavamian, Maria Haupt, and Steve Reynolds for useful discussions and Annie Hughes for sharing the ATCA and Parkes survey data of the LMC. Additionally, we thank Daniel Wik for assistance with {\it nuskybgd}. We also acknowledge the continued work of the astronomical community contributing to AstroLib. Our work was supported under NASA contract NNG08FD60C and made use of data from the {\it NuSTAR} mission, a project led by the California Institute of Technology, managed by the Jet Propulsion Laboratory, and funded by NASA. We thank the {\it NuSTAR} Operations, Software and Calibration teams for support with the execution and analysis of these observations. This research made use of the {\it NuSTAR} Data Analysis Software (NuSTARDAS), jointly developed by the ASI Science Data Center (ASDC, Italy) and the California Institute of Technology (USA). Support for this work was provided by National Aeronautics and Space Administration through Chandra Award Number G08--19062X and through Smithsonian Astrophysical Observatory contract SV3--73016 to MIT issued by the Chandra X-ray Observatory Center, which is operated by the Smithsonian Astrophysical Observatory for and on behalf of NASA under contract NAS8--03060.

\software{CIAO (v4.7; \citealt{fru06}), XSPEC (v12.9.0; \citealt{arnaud96}), ftools \citep{blackburn95}, NuSTAR Data Analysis Software (v1.8.0), {\it XMM-Newton} SAS (v15.0.0; \citealt{gabriel04})}

\nocite{*}
\bibliographystyle{aasjournal}
\bibliography{30dorc}

\end{document}